\documentclass[reprint,secnumarabic,amssymb, nobibnotes, aps, prl,superscriptaddress,showpacs]{revtex4-1}

\usepackage{amsmath}
\usepackage{amssymb}
\usepackage{booktabs}
\usepackage{float}
\usepackage{graphics}
\usepackage{graphicx}
\usepackage{hyperref}
\usepackage{here}

\def\gb{$\bar{\Gamma}$\ }
\def\mb{$\bar{\mathrm{M}}$\ }
\def\mbp{$\bar{\mathrm{M}}$.\ }
\def\gmb{${\bar{\Gamma}}{\bar{\mathrm{M}}}$\ }
\def\EF{$E_{\mathrm{F}}$\ }
\def\EFp{$E_{\mathrm{F}}$.\ }

\begin{document}

\title{Proving Nontrivial Topology of Pure Bismuth by Quantum Confinement}
\author{S. Ito}
\affiliation{Institute for Solid State Physics (ISSP), The University of Tokyo, Kashiwa, Chiba 277-8581, Japan}
\author{B. Feng}
\affiliation{Institute for Solid State Physics (ISSP), The University of Tokyo, Kashiwa, Chiba 277-8581, Japan}
\author{M. Arita}
\affiliation{Hiroshima Synchrotron Radiation Center (HSRC), Hiroshima University, Higashi-Hiroshima, Hiroshima 739-0046, Japan}
\author{A. Takayama}
\affiliation{Department of Physics, The University of Tokyo, Bunkyo-ku, Tokyo 113-0033, Japan}
\author{R.-Y. Liu}
\affiliation{Institute for Solid State Physics (ISSP), The University of Tokyo, Kashiwa, Chiba 277-8581, Japan}
\author{T. Someya}
\affiliation{Institute for Solid State Physics (ISSP), The University of Tokyo, Kashiwa, Chiba 277-8581, Japan}
\author{W.-C. Chen}
\affiliation{National Synchrotron Radiation Research Center (NSRRC), Hsinchu, Taiwan 30076, Republic of China}
\author{T. Iimori}
\affiliation{Institute for Solid State Physics (ISSP), The University of Tokyo, Kashiwa, Chiba 277-8581, Japan}
\author{H. Namatame}
\affiliation{Hiroshima Synchrotron Radiation Center (HSRC), Hiroshima University, Higashi-Hiroshima, Hiroshima 739-0046, Japan}
\author{M. Taniguchi}
\affiliation{Hiroshima Synchrotron Radiation Center (HSRC), Hiroshima University, Higashi-Hiroshima, Hiroshima 739-0046, Japan}
\author{C.-M. Cheng}
\affiliation{National Synchrotron Radiation Research Center (NSRRC), Hsinchu, Taiwan 30076, Republic of China}
\author{S.-J. Tang}
\affiliation{National Synchrotron Radiation Research Center (NSRRC), Hsinchu, Taiwan 30076, Republic of China}
\affiliation{Department of Physics and Astronomy, National Tsing Hua University, Hsinchu, Taiwan 30013, Republic of China}
\author{F. Komori}
\affiliation{Institute for Solid State Physics (ISSP), The University of Tokyo, Kashiwa, Chiba 277-8581, Japan}
\author{K. Kobayashi}
\affiliation{Department of Physics, Ochanomizu University, Bunkyo-ku, Tokyo 112-8610, Japan}
\author{T.-C. Chiang}
\affiliation{Department of Physics and Frederick Seitz Materials Research Laboratory, University of Illinois at Urbana-Champaign, Urbana, Illinois 61801, USA}
\author{I. Matsuda}
\affiliation{Institute for Solid State Physics (ISSP), The University of Tokyo, Kashiwa, Chiba 277-8581, Japan}
\date{\today}
\begin{abstract}
The topology of pure Bi is controversial because of its very small ($\sim$10 meV) band gap. Here we perform high-resolution angle-resolved photoelectron spectroscopy measurements systematically on 14$-$202 bilayer Bi films. Using high-quality films, we succeed in observing quantized bulk bands with energy separations down to $\sim$10 meV. Detailed analyses on the phase shift of the confined wave functions precisely determine the surface and bulk electronic structures, which unambiguously show nontrivial topology. The present results not only prove the fundamental property of Bi but also introduce a capability of the quantum-confinement approach.
\end{abstract}
\maketitle

Semimetal bismuth (Bi) has been providing an irreplaceable playground in condensed matter physics. Its extreme properties originating from the three-dimensional Dirac dispersion enabled the first observations of several important phenomena such as diamagnetism \cite{brugmans1778} and the various effects associated with Seebeck \cite{seebeck1826}, Ettingshausen and Nernst \cite{ettingshausen1886}, Shubnikov and de Haas \cite{schubnikov1930} and de Haas and van Alphen \cite{dehaas1930}. Even now, numbers of novel quantum phenomena have been intensively reported on this system \cite{behnia2007,li2008,fuseya2009,zhu2012,kuchler2014,fuseya2015,collaudin2015,du2016}. In spite of the enormous amount of research, one fundamental property of Bi has been controversial: its electronic topology. Because of its huge spin-orbit coupling (SOC) \cite{hofmann2006}, Bi has also been a central element in designing topological materials such as Bi$_{1-x}$Sb$_x$, Bi$_2$Se$_3$, Na$_3$Bi, and $\beta$-Bi$_4$I$_4$ \cite{hasan2010,qi2011,wang2012,autes2015,liu2016}. A combination of SOC and several symmetries produces topologically protected electronic states with inherent spin splitting. Despite the essential role in topological studies, a pure Bi crystal itself had long been believed topologically trivial based on several calculations \cite{golin1968,liu1995,koroteev2004,koroteev2008,teo2008, hirahara2012,aguilera2015}, which had been considered to agree with transport \cite{red1985} and angle-resolved photoelectron spectroscopy (ARPES) measurements \cite{ast2001,ast2003,koroteev2004}. However, a recent high-resolution ARPES result suggests the surface bands are actually different from previously calculated ones and Bi possesses a nontrivial topology \cite{ohtsubo2013,perfetti2015}. New transport measurements also imply the presence of topologically protected surface states \cite{ning2014_2,ning2014}.

\begin{figure}[b]
\includegraphics{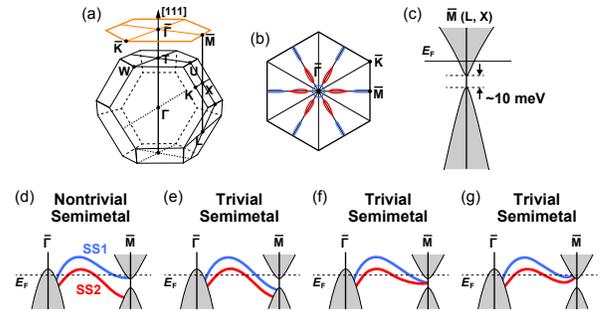}
\caption{\label{fig1} Schematic representation of (a) the bulk and surface Brillouin zone of Bi crystal in the [111] direction and (b) the Fermi surface. (c) Near-\EF structure of the bulk projections at \mb. (d)$-$(g) Possible band structures along the \gmb direction on the Bi(111) surface. The blue and red lines indicate the two spin-splitting surface bands, SS1 and SS2, respectively.}
\end{figure}

Nevertheless, the recent ARPES result has not yet been conclusive because it lacks clear peaks of bulk bands \cite{ohtsubo2013,perfetti2015}. In principle, surface-normal bulk dispersions can be measured by changing the incident photon energy, where the momentum resolution is determined from the uncertainty relation $\Delta z\cdot \Delta k_z \geq 1/2$ (Ref. \cite{hufner2013}). ($\Delta z$ is an escape depth of photoelectrons.) However, the Dirac dispersion of Bi is so sharp against this resolution that $h\nu$-dependent spectra show no clear peak \cite{ast2003,ohtsubo2013,perfetti2015}. This is a serious problem because Bi has a very small ($\sim$10 meV \cite{liu1995,aguilera2015}) band gap and a slight energy shift in bulk bands can easily transform a nontrivial case [Fig. \ref{fig1}(d)] into a trivial case [Fig. \ref{fig1}(e)]. In short, to unambiguously identify the topology of Bi, one must precisely determine both the surface and bulk electronic structures.
One promising approach is using a thin film geometry, where quantum-well state (QWS) subbands are formed inside bulk band projections \cite{chiang2000,hirahara2006}. Although QWSs originate from bulk states, they possess a two-dimensional character and can be clearly observed in ARPES measurements. 
%However, the smallest QWS separations previously observed were $\sim$0.1 eV \cite{hirahara2007_1}, which is still $\sim$10 times larger than Bi band gap.

In this Letter, we performed high-resolution ARPES measurements on  Bi(111) films with thicknesses increasing from 14 to 202 BL (bilayer; 1 BL = 3.93 \AA \ \cite{liu1995}). High-quality films enabled us to clearly observe the QWS subbands with energy separations down to $\sim$10 meV. After we confirmed the interaction between the top and bottom surface states in the 14 BL film, we systematically followed the evolution of the electronic structures. Detailed analyses on the phase shift of the QWS wave functions precisely determined the surface and bulk band dispersions. The revealed electronic structures unambiguously show that a pure Bi crystal has a nontrivial topology. The present results not only prove the fundamental property of Bi, but also highlight the QWS approach as a powerful tool to determine fine electronic structures.

A surface of a \textit{p}-type Ge wafer cut in the [111] direction was cleaned in ultrahigh vacuum by several cycles of Ar$^+$ bombardment and annealing up to 900 K. Bi was deposited at room temperature and annealed at 400 K \cite{hatta2009}. The pressure was kept at $\sim$1$\times10^{-8}$ Pa during the deposition. The film thickness was precisely measured with a quartz thickness monitor. The qualities of the substrate and the film were confirmed from low-energy electron diffraction measurements. ARPES measurements were performed at BL-9A of HSRC and BL-21B1 of NSRRC. In BL-9A a high-intensity unpolarized Xe plasma discharge lamp (8.437 eV) was used in addition to synchrotron radiation (21 eV). The measurement temperature was kept at 10 K, and the total energy resolution was 12 meV for 21 eV photons and 7 meV for 8.437 eV photons. The first-principles calculations were performed using the VASP computer code \cite{kresse1996}. A free-standing slab was used based on previous reports \cite{hirahara2006,koroteev2008,takayama2012}. (See the Supplemental Material \footnote{See Supplemental Material at [url] for details of the first-principles calculations, calcualtions of $E$-$k_\perp$ dispersion, and peak fittings, which includes Refs. [21,23,38,41-44].}, which includes Refs. \cite{liu1995,koroteev2008,kresse1996,perdew1996, blochl1994, takayama2015, norman1998}.)

First we organize information regarding the Bi topology. For the (111) surface of Bi, two spin-splitting surface bands SS1 and SS2 bridge the \gb and \mb points. Although experimental and theoretical results agree that both bands connect to the valence band (VB) around the \gb point, a discrepancy lies in their connection around the \mb point \cite{ast2001,ast2003,koroteev2004,hofmann2006,ohtsubo2013}. Based on Kramers's theorem, a spin-splitting band cannot exist at time-reversal-invariant momenta (TRIM) \cite{hasan2010,qi2011}. Therefore, we can limit the possible cases to those depicted in Figs. \ref{fig1}(d)$-$1(g). We note a nontrivial topology exists only in the Fig. \ref{fig1}(d) case, which is distinguished from the other cases in that the SS1 and SS2 bands are nondegenerate at \mbp

\begin{figure}
\includegraphics{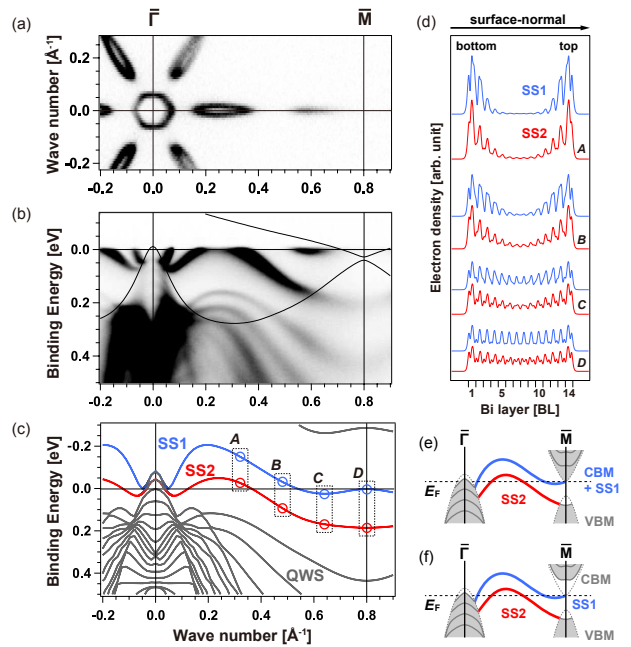}
\caption{\label{fig2}(a), (b) The Fermi surface and the band structures measured along the \gmb direction in a 14 BL Bi(111) film at $h\nu=$ 21 eV. Solid lines in (b) indicate bulk projections calculated by a tight-binding method \cite{liu1995}. (c) Band structures obtained by the first-principles calculations for a 14 BL Bi slab. (d) Plane-averaged electron densities within the film calculated at the four $k$ points marked in (c).  (e), (f) Possible band assignments in an ultrathin Bi film. Gray areas illustrate positions of the VB maximum (VBM) and CB minimum (CBM).}	
\end{figure}

We start from an observation of an ultrathin film. Figure \ref{fig2}(a) shows the Fermi surface of a 14 BL Bi(111) film measured at $h\nu=$ 21 eV. The shape is very close to that of bulk Bi \cite{ast2001,ohtsubo2012,ohtsubo2013}. Figure \ref{fig2}(b) shows the corresponding band structures along the \gmb direction with calculated bulk projections. Two surface bands exist inside the bulk band gap and QWS subbands inside the bulk projection. The observed bands are consistent with previous reports \cite{hirahara2006,hirahara2007_1,bian2009}. Figure \ref{fig2}(c) illustrates the band structures obtained by the first-principles calculations. Although there is a slight discrepancy in the energy positions, the overall structures show good qualitative agreement.

It is clear that the SS1 and SS2 bands are non-degenerate at \mb, which appears to suggest that Bi is topologically nontrivial based on Figs. \ref{fig1}(d)$-$(g). However, in an ultrathin Bi film whose thickness is as small as a decay length of the surface state, the top and bottom surface states can interact with each other and modify their shape from the bulk limit \cite{hirahara2006, takayama2012, hirahara2015}. Figure \ref{fig2}(d) shows plane-averaged electronic charge densities within the film calculated at the four $k$ points marked in Fig. \ref{fig2}(c). Although these states are actually localized on surfaces near the center of the Brillouin zone (\textit{A, B}), they gradually penetrate into the film and form bulklike states in approaching \mb point (\textit{C, D}). Because the state \textit{C} lies far from bulk projections around \mb, this bulklike behavior arises indeed from such a surface-surface interaction. These merged states possess even numbers of electrons and can exist inside a band gap at TRIM without violating Kramers's theorem. Therefore, in addition to the nontrivial scenario that SS1 connects to the conduction band (CB) at \mb [Fig. \ref{fig2}(e)], it is also possible that SS1 connects to the VB in the bulk limit but that it is pushed into a gap in an ultrathin film by the surface-surface interaction [Fig. \ref{fig2}(f)] \cite{hirahara2006, ohtsubo2013}. Although Figs. \ref{fig2}(e) and 2(f) depict SS2 hybridizing with the VB at \mb as suggested by previous studies \cite{koroteev2004,ohtsubo2013}, it must also be tested. To identify Bi topology, we have to follow the evolution of SS1 and SS2. If they never cross each other even in the bulk limit, there is no choice but the Fig. \ref{fig2}(e) [that is, Fig. \ref{fig1}(d))] case, which unambiguously proves pure Bi is topologically nontrivial. 

\begin{figure}[b]
\includegraphics{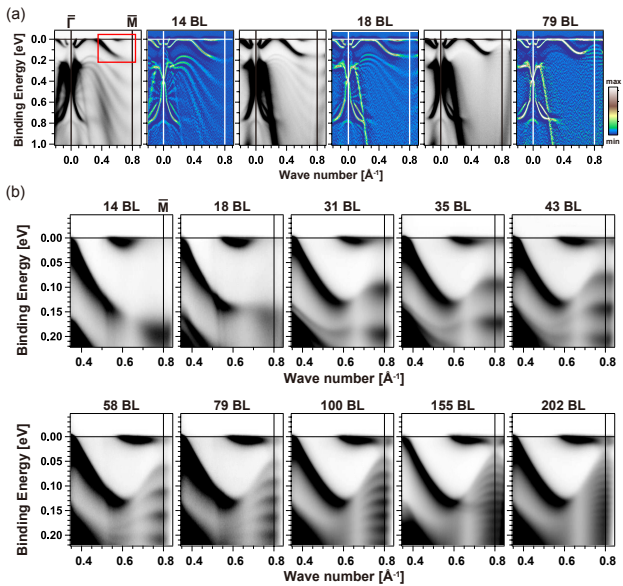}
\caption{\label{fig3}(a) Wide-range band structures measured along the \gmb direction in 14, 18, and 79 BL Bi(111) films at $h\nu=$ 21 eV. The colored images were produced using a curvature method for better visualization \cite{zhang2011}. (b) Near-$E_{\mathrm{F}}$ band structures measured at $h\nu=$ 8.437 eV inside the red box in (a). The thickness is systematically increased from 14 to 202 BL.}
\end{figure}

Figure \ref{fig3}(a) shows the wide-range band structures measured along the \gmb direction at $h\nu=$ 21 eV for 14, 18, and 79 BL films. Whereas quantized bands were clearly observed in 14 and 18 BL films, these bands became almost continuous in a 79 BL film except for a region near \EF around \mb. To observe the area in more detail, we performed ARPES measurements with higher energy resolution at $h\nu=$ 8.437 eV. Figure \ref{fig3}(b) shows ARPES images taken inside the red box in Fig. \ref{fig3}(a). The thicknesses of the films are systematically increased from 14 to 202 BL. As the thickness increases, a QWS energy separation decreases from $\sim$200 to $\sim$10 meV. A series of QWS subbands near \mb gradually converges into the projected VB and the intensity of the SS2 band drops abruptly when it crosses the edge. This implies SS2 around \mb strongly hybridizes with bulk states and becomes a part of the QWSs.

\begin{figure}
\includegraphics{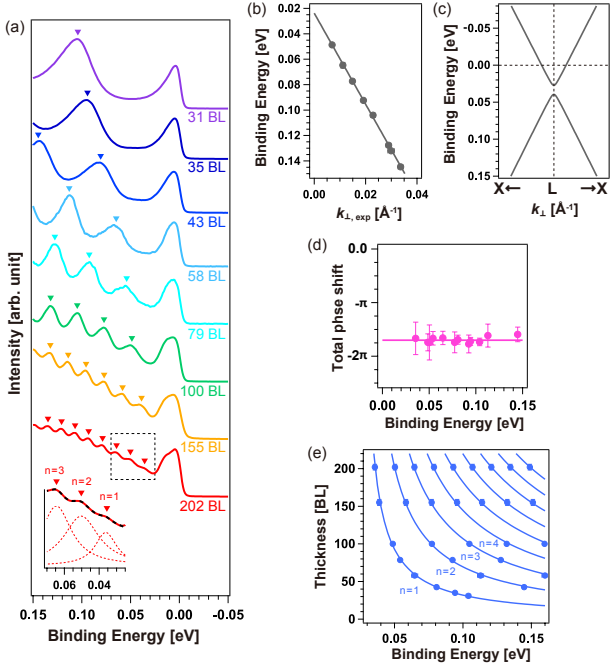}
\caption{\label{fig4} (a) EDCs extracted at \mb ($k=0.8$ \AA$^{-1}$). The triangles show peak positions fitted by Lorentzian functions [inset]. (b) $E$-$k_\perp$ dispersion experimentally obtained using Eq. (\ref{eq2}). The solid line represents a linear fit. (c) $E$-$k_\perp$ dispersion obtained from a tight-binding calculation \cite{liu1995}. (d) Total phase shifts experimentally derived using Eq. (\ref{eq1}). (e) A plot of the $N$-$E$ relation in QWSs (a structure plot). Solid lines are drawn using Eq. (\ref{eq3}).}	
\end{figure}

To test this hypothesis, we analyzed the QWS energy positions in more detail. Figure \ref{fig4}(a) shows the energy distribution curves (EDCs) extracted at \mb for each thickness. Peak positions were determined using Lorentzian fittings. These energy positions can be simply described using the phase accumulation model, which assumes electronic waves propagating forward and backward across the film and being reflected at the top and bottom surfaces \cite{chiang2000}. The model provides the expression
\begin{equation}
	\label{eq1}
	2k_\perp(E)N(E)t+\Phi(E) = 2\pi(n-1)
\end{equation}
The first term represents the phase shifts in propagation, with $k_\perp(E)$ and $N(E)$ denoting the surface-normal dispersion and the number of bilayers, respectively, and $t$ the thickness of one bilayer (3.93 \AA \ \cite{liu1995}); $\Phi(E)$ is the total phase shift at the top and bottom surfaces and $n$ is a quantization number. 

To experimentally extract information concerning $k_\perp$, we note that some QWSs have the same binding energy but different $N$ and $n$. Since the phase shift $\Phi$ can be regarded as only a function of $E$ \cite{chiang2000}, we can derive 
\begin{equation}
	\label{eq2}
	k_{\perp, \mathrm{exp}}=\frac{\pi}{t}\frac{n-n'}{N-N'}
\end{equation}
Figure \ref{fig4}(b) shows the $E$-$k_{\perp, \mathrm{exp}}$ dispersion obtained using this relation \cite{Note1}. The error bars are estimated by uncertainties in thicknesses and fitted peak positions. Here the surface-normal direction at \mb corresponds to LX [Fig. \ref{fig1}(a)] and Bi has its Dirac dispersion along this direction. Figure \ref{fig4}(c) shows the tight-binding result \cite{liu1995}. The experimental data are indeed perfectly fitted by the solid line in Fig. \ref{fig4}(b); the fitted result is $E=\alpha k_{\perp, \mathrm{exp}}+\beta$, where $\alpha=3.58 \pm 0.11$ eV$\cdot$\AA \ and $\beta=0.024 \pm 0.002$ eV. 

Now that we have experimentally obtained $k_\perp(E)$, we can derive a total phase shift using Eq. (\ref{eq1}). For this purpose, we used $n=1$ and $n=2$ QWS energy positions and corresponding thicknesses. The result shown in Fig. \ref{fig4}(d) exhibits an almost constant relation in this energy range. The fitted value by a constant function is $\Phi_{\mathrm{exp}}=(-1.70\pm0.03)\pi$, which is similar to those reported in ultrathin Bi films on a Si substrate \cite{hirahara2007_1}. Furthermore, we compared the experimental and analytical results by plotting $N$ against $E$ (a structure plot) in Fig. \ref{fig4}(e). The latter is obtained using 
\begin{equation}
	\label{eq3}
	N(E)=\frac{2\pi(n-1)-\Phi_{\mathrm{exp}}}{2k_{\perp, \mathrm{exp}}(E)t}
\end{equation}
It excellently reproduces the experimental data not only for $n=1$ and $n=2$ QWSs but also for each of the other $n$ values. The consistency of the entire analysis shows that SS2 band around \mb indeed becomes a part of QWSs, and also demonstrates the validity of the obtained phase shift.

\begin{figure}
\includegraphics{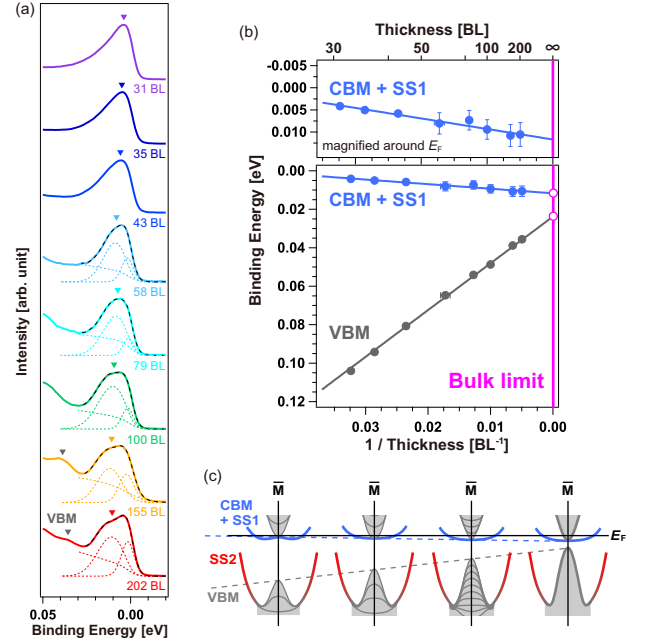}
\caption{\label{fig5} (a) EDCs at \mb magnified around \EFp (b) Evolutions of peak positions extracted in (a) and those of VBM ($n=1$ QWS in VB) against an inverse thickness $1/N$. The meanings of the solid lines are discussed in the text. (c) Schematic representation of the evolution in electronic structures of Bi films approaching the bulk limit.}	
\end{figure}

As a final step we follow the evolution of the VB and SS1 bands at \mb to identify Bi topology. Figure \ref{fig5}(a) shows EDCs magnified around a peak near \EFp The peak broadens as thickness increases and finally exhibits multiple peaks. This is attributed to a tail of a QWS located above \EFp We noted the clear threshold between 43 and 58 BL films and applied a specific fitting method for films above 43 BL \cite{Note1}. Extracted peak positions were plotted against an inverse thickness $1/N$ along with the VBM ($n=1$ QWS) peaks in Fig. \ref{fig5}(b). Using Eq. (\ref{eq1}), an inverse thickness $1/N$ and a surface-normal wave number $k_\perp$ are simply connected by $k_\perp=-\Phi/2Nt$ at the VB and CB edges ($n=1$). Since the total phase shift turns out to be constant within this energy range, the VBM evolution is expressed as
\begin{equation}
	\label{eq4}
	E=-\frac{\alpha \Phi_{\mathrm{exp}}}{2t}\frac{1}{N}+\beta
\end{equation}
The gray solid line in Fig. \ref{fig5}(b) represents this linear function, which perfectly reproduces the experimental data.

The evolution of the SS1 peak also appears to fit a linear function, suggesting a hybridization between the CBM and SS1. To test it, we extended the phase analysis for the VB to CB. A simple two-band model indicates that a total phase shift of a QWS wave function is strongly affected by the parity and changes its value by $2\pi$ across the band gap \cite{smith1985}. The blue solid line in Fig. \ref{fig5}(b) is a linear fit, whose gradient can be reproduced by Eq. (\ref{eq4}) when $\Phi_{\mathrm{CBM}}=\Phi_{\mathrm{VBM}}+1.87\pi$. Here we used the same $\alpha$ value as for the VB based on completely symmetric dispersions shown in Fig. \ref{fig4}(c). The close correspondence with $2\pi$ strongly suggests that the peak near \EF belongs to a QWS at the CBM that directly hybridizes with SS1.The CBM and VBM values in the bulk limit are 0.012 $\pm$ 0.002 eV and 0.024 $\pm$ 0.002 eV, respectively, which results in a gap of 0.012 $\pm$ 0.003 eV. It is quite consistent with previous reports \cite{liu1995,aguilera2015}. The fact that the CBM (SS1) and VBM never cross even in the bulk limit excludes all possibilities but that of Fig. \ref{fig1}(d), a nontrivial semimetal.

One may be concerned about the CBM and VBM positions deviating by $\sim$0.015 eV from previous values \cite{liu1995,aguilera2015} [e.g., Fig. \ref{fig4}(c)]. A possible reason is a strain effect from the Ge substrate. However, this can be excluded by considering the $1/N$ dependence. A lattice strain exhibits an exponential decay against the film thickness \cite{yao2016}, but the linear dispersion in Fig. \ref{fig5}(b) does not appear to fit an exponential decay. Moreover, an exponential function has downward convexity with $1/N$, which further reduces the possibility that the VBM and CBM cross each other. 

In conclusion, we were able to unambiguously prove that pure Bi is topologically nontrivial. Although the interaction between the top and bottom surface states does exist as revealed by calculations, the splitting between SS1 and SS2 is not a consequence of the interaction but rather the electronic structure unique to Bi. The present result provides an important insight for recent attempts to detect novel quantum phenomena on pure Bi, where the three-dimensional massive Dirac fermion and its nontrivial topology can show an interesting connection. Furthermore, the topologically protected surface states with a giant spin splitting offer great potential in spintronics applications. Recent transport measurements have shown Bi keeps its unique surface transport at ambient pressure \cite{ning2014,ning2014_2}. A possible application of Bi surface states to valleytronics was also recently reported \cite{du2016}. 

Finally we also emphasize the capability of the QWS approach we used. 
Further advancing the established method \cite{chiang2000,hirahara2007_1}, we demonstrated that systematic analyses on QWSs can precisely assign and map surface and bulk bands even at $\sim$10 meV scale and can reveal hybridizations between them.
Novel topological materials recently predicted can have as small energy scales as observed here in Bi \cite{hirayama2015, ruan2016}. Precise determination of surface and bulk electronic structures is indispensable in driving forward topological studies, where the present method can be one of the most powerful tools.

\begin{acknowledgements}
	We acknowledge Y. Ohtsubo, K. Yaji, K. Kuroda Y. Ishida, S. Yamamoto, and P. Zhang for valuable discussions. We also thank C.-H. Lin, D.-X. Lin and F.-Z. Xiao for their experimental help. The ARPES measurements were performed with the approval of the Proposal Assessing Committee of HSRC (Proposal No. 15-A-38) and the Proposal Assessing Committee of NSRRC (Project No. 2015-2-090-1). T.-C.C. acknowledges support by the U.S. National Science Foundation under Grant No. DMR-1305583. S.I. was supported by JSPS through Program for Leading Graduate Schools (ALPS).
\end{acknowledgements}

\bibliography{reference.bib}

\end{document}